\documentclass[conference,letterpaper]{IEEEtran}
\IEEEoverridecommandlockouts
\usepackage{cite}
\usepackage{amsmath,amssymb,amsfonts, amsthm}
\usepackage{algorithmic}
\usepackage{algorithm}
\usepackage{graphicx}
\usepackage{textcomp}
\usepackage{xcolor}

\usepackage{mathtools}
\def\BibTeX{{\rm B\kern-.05em{\sc i\kern-.025em b}\kern-.08em
    T\kern-.1667em\lower.7ex\hbox{E}\kern-.125emX}}
\newtheorem{theorem}{Theorem}
\newtheorem{lemma}{Lemma}
\newtheorem{observation}{Observation}
\newtheorem{claim}{Claim}
\newtheorem{corollary}{Corollary}

\theoremstyle{remark}
\newtheorem*{remark}{Remark}

\makeatletter
\def\thm@space@setup{%
  \thm@preskip=0pt
  \thm@postskip=0pt
}

\def\th@remark{%
  \normalfont 
  \thm@preskip=0pt 
  \thm@postskip=0pt 
}
\makeatother

\usepackage{geometry}
\newcommand{\capsymbol}{\mathsf{cap}}

\newcommand{\costsymbol}{\mathsf{cost}}

\begin{document}

\title{Studying the Cycle Complexity of DNA Synthesis\vspace{-1ex}
{\footnotesize \textsuperscript{}}}
\author{\IEEEauthorblockN{\textbf{Amit Zrihan}\IEEEauthorrefmark{1}, \textbf{Eitan Yaakobi}\IEEEauthorrefmark{1}, and \textbf{Zohar Yakhini}\IEEEauthorrefmark{1}\IEEEauthorrefmark{2}}
\IEEEauthorblockA{\IEEEauthorrefmark{1}Faculty of Computer Science, Technion - Israel Institute of Technology, Israel}
\IEEEauthorblockA{\IEEEauthorrefmark{2}School of Computer Science, RUNI, Israel}
\IEEEauthorblockA{\{amit.zr@campus.technion.ac.il, yaakobi@cs.technion.ac.il, zohar.yakhini@gmail.com\}\vspace{-5ex}}
\thanks{This research was funded by the European Union (DiDAX, 101115134). Views and opinions expressed are however those of the author(s) only and do not necessarily reflect those of the European Union or the European Research Council Executive Agency. Neither the European Union nor the granting authority can be held responsible for them.}}
\vspace{-5ex}\maketitle
\begin{abstract}
Storing data in DNA is being explored as an efficient solution for archiving and in-object storage.
Synthesis time and cost remain challenging, significantly limiting some applications at this stage. In this paper we investigate efficient synthesis, as it relates to cyclic synchronized synthesis technologies, such as  photolithography. We define performance metrics related to the number of cycles needed for the synthesis of any fixed number of bits.
We first expand on some results from the literature related to the channel capacity, addressing densities beyond those covered by prior work. This leads us to develop effective encoding achieving rate and capacity that are higher than previously reported. Finally, we analyze cost based on a parametric definition and determine some bounds and asymptotics. We investigate alphabet sizes that can be larger than 4, both for theoretical completeness and since practical approaches to such schemes were recently suggested and tested in the literature.
\end{abstract}

\renewcommand{\baselinestretch}{0.86}\normalsize
\section{Introduction}
Digital data can be stored in DNA oligonucleotides thanks to the composition of DNA molecules, which consist of four nucleotides: Adenine (A), Cytosine (C), Guanine (G), and Thymine (T).
An oligonucleotide (oligo), represents a sequence of these nucleotides in a specific order, and can be denoted as a string over the alphabet $\{A, C, G, T\}$.
As it is feasible to chemically synthesize nearly any nucleotide sequence, DNA oligos can serve to store digital data.

The initial phase of DNA data storage involves converting binary data into DNA oligos.
This process involves adding redundant bases for indexing and using coding techniques to reduce errors and losses that commonly occur during synthesis and sequencing.
The next step consists of synthesis.
In this step each oligo in the storage system is typically synthesized thousands to millions of times, resulting in a vast collection of unordered DNA fragments.
Storing the DNA which now codes the information may require the use of technology to guarantee its long time integrity \cite{Grass2015RobustCP}.
During the reading process, a subset of these fragments is then selected and read using sequencing methods.
To reconstruct the stored sequences, the resulting reads are grouped into clusters that are likely originated from the same design oligo.
The original designed oligos are then reconstructed using the reads (see e.g \cite{Shafir2024SequenceDA,Shafir2021SequenceRU}).
The original data is then retrieved from the reconstructed DNA sequences \cite{Church, Goldman, Grass_2015, BLAWAT20161011}.

Photolithographic DNA synthesis \cite{Antkowiak2020LowCD, Schaudy2023EnzymaticSO, behr2024} is a technique used in the synthesis step.
Photolithography can also be used in other synthesis applications, including for direct synthesis of RNA \cite{Lietard2018HighDensityRM}.
In the context of DNA for data storage, the process results in a large number of DNA oligonucleotides, all synthesized in parallel. Under the standard approach, each oligo is grown by at most one nucleotide in each synthesis cycle.
The machine follows a fixed supersequence of possible nucleotides to append to the oligos.
As the machine iterates through this supersequence, the next nucleotide is added to a selected subset of the oligos until the machine reaches the end of the supersequence.
In particular, each synthesized DNA oligo must be a subsequence of the machine’s supersequence.
In this paper, we operate under the assumption that there are no constraints on the number of oligos that can be synthesized simultaneously for any practical run of the machine.

Several approaches were suggested and implemented, however, that support larger alphabet
\cite{US-20210141568-A1, US-20230040158-A1, US-20240013063-A1, Anavy2019DataSI, Preuss2024, Choi2019HighIC, Yan2023}.
Throughout the paper, we will work with the abstract alphabet $\Sigma_q \triangleq \{1,2,\ldots,q\}$,
where each $\sigma \in \Sigma_q$ is referred to as a symbol.
As the synthesis machine's supersequence we will use the alternating sequence $A_q$, that cyclically repeats all symbols in $\Sigma_q$ in ascending order.
For an alphabet of size $q$, the alternating sequence is $A_q = (123\cdots q123\cdots q\cdots)$.
For a given number of synthesis cycles $C$ we will use $A_q[C]$ to denote the length-$C$ prefix of $A_q$ as the supersequence.
When synthesizing a predetermined length of oligonucleotides, let $L$ denote that length, and denote also $\rho \triangleq \frac{L}{C}$ $(0\leq \rho \leq 1)$.
One of our main objectives in this paper is to minimize the number of synthesis cycles that is required in order to store some $N$ information bits.
Hence, our aim is to maximize, for a single oligo, the ratio between the number of information bits and the number of synthesis cycles.
This is in the spirit of the term \emph{logical density} used in some of the literature.
Another main objective to this work is to investigate the cost of synthesis under various assumptions.
These questions will be precisely defined in the next section. The rest of the paper is structured as follows.
In Section~\ref{section:PrelimAndFormulations},  we provide some technical preliminaries and definitions.
In Section~\ref{section:Encoders}, we address encoders that attain previously proven bounds \cite{Lenz2021MultivariateAC}.
In Section~\ref{section:CostOptimization}, we analyze the synthesis cost.
Finally, a discussion of limitations and future work is given in Section~\ref{section:Discussion}.
Due to space constraints, some of the proofs are omitted.

\section{Preliminaries and problem formulation}\label{section:PrelimAndFormulations}
\subsection{Information Theory Limits of Photolithographic Synthesis}
\begin{figure}[htbp]
    \centerline{\includegraphics[width=7.5cm,height=4.5cm]{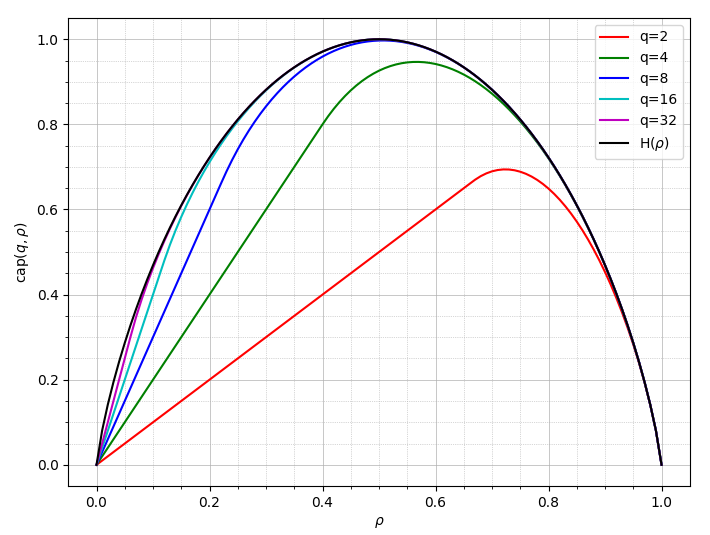}}
    \setlength{\abovecaptionskip}{-5pt}
    \caption{$\capsymbol(q,\rho)$ as a function of $q$ and $\rho$.}
    \label{fig:Capacity_fixed_length}
\end{figure}

The \emph{radius-t deletion sphere} (exactly $t$ deletions) of a string $\mathbf{s}$ is denoted by $D(\mathbf{s},t)$, and $D_q(C,t) \triangleq |D(A_q[C],t)|$. 
As indicated in \cite{Hirschberg2000TightBO} the recursive formula $D_q(C,t) = \sum_{i=0}^{t}\binom{C-t}{i}D_{q-1}(t, t-i)$ holds.
Let $M_q(C,L)$ be the number of distinct subsequences of length $L$ of $A_q[C]$, and note that $M_q(C,L) = D_q(C,C-L)$.
\begin{lemma}\label{lemma:MqUpperBound}
    For any $q,C,L \in \mathbb{N}, (L\leq C)$ it holds that
    \begin{equation}\label{eq:Mq}
    M_q(C,L) \leq M_{q+1}(C,L) \leq \binom{C}{L},
    \end{equation}
    where the inequality is strict for $2 \leq q < C, 2 \leq C, 0 < L$.
\end{lemma}
\begin{proof}
    The first inequality can be shown by induction on $q$ and the recursive formula of $D_q(C,t)$.
    For the second inequality, $\binom{C}{L}$ represents the number of ways to choose $L$ indices out of string of length $C$.
    However, since these options may not necessarily generate distinct strings, the inequality arises.
\end{proof}
\begin{lemma}\label{lemma:MqLowerBound}
    It holds that $\binom{v}{u}^n \leq M_v(n \cdot v, n \cdot u)$, $(u,v,n \in \mathbb{N})$.
\end{lemma}
\begin{proof}
    Consider a scenario where for each window of length $v$ we select $u$ symbols for the subsequence, yielding a subsequence of length $n\cdot u$.
    Every distinct choice made in every window results in a distinct subsequence.
    Consequently, the total number of ways to choose $u$ symbols out of $v$, $n$ times is $\binom{v}{u}^n$.
    However, since this is represent only a subset of all the possible subsequences, the inequality holds.
\end{proof}
We define, analogous to~\cite{Lenz2021MultivariateAC}
\begin{align*}
    &\capsymbol(q,\rho) \triangleq \limsup_{C\rightarrow \infty}{\frac{\log_2{M_q(C,\lfloor \rho \cdot C \rfloor)}}{C}},\\
    &\capsymbol(q) \triangleq \limsup_{C\rightarrow \infty}{\frac{\log_2{\left(\sum_{L=0}^C M_q(C,L)\right)}}{C}}.
\end{align*}
These values represent the maximum number of information bits that can be stored per synthesis cycle for a fixed oligonucleotide length $L = \lfloor \rho\cdot C \rfloor$ and for flexible $L$, respectively. All when synthesizing a single oligo under $A_q$ as the supersequence of length $C$.
\begin{theorem}
    It holds that
    \begin{align}
    &\capsymbol(q,\rho) \leq \capsymbol(q+1, \rho) \leq \lim\limits_{q\to \infty} \capsymbol(q,\rho) = H(\rho)\label{eq:CapqEntropy}\\
    &\capsymbol(q) \leq \capsymbol(q+1) \leq \lim\limits_{q\to \infty}\capsymbol(q) = 1\label{eq:Cap1}
    \end{align}
    where $H$ is the binary entropy function.
\end{theorem}
\begin{proof}
    As for~(\ref{eq:CapqEntropy}), both inequalities follows from Lemma~\ref{lemma:MqUpperBound}.
    As for the the limit, we start from an upper bound
    \begin{align*}
        \lim_{q \to \infty}\capsymbol(q&,\rho) = \lim_{q \to \infty} \limsup_{C\rightarrow \infty}{\frac{\log_2{M_q(C,\lfloor \rho \cdot C \rfloor)}}{C}}\\
        &\underset{\mathrm{Lemma~\ref{lemma:MqUpperBound}}}{\leq} \lim_{q \to \infty} \limsup_{C\rightarrow \infty}{\frac{\log_2{\binom{C}{\lfloor \rho \cdot C \rfloor}}}{C}} = H(\rho).
    \end{align*}
    We showed that the sequence $\capsymbol(q,\rho)$ is monotonously increasing and bounded from above therefore it converges.
    As for the lower bound, let $\rho = \frac{s}{t} \in \mathbb{Q}$.
    It holds that
    \begin{align*}
        \lim_{q \to \infty} \capsymbol(q,\rho) &= \lim_{n \to \infty} \capsymbol(n\cdot t,\rho)\\ 
        &= \lim_{n \to \infty} \limsup_{C\rightarrow \infty}{\frac{\log_2{M_{n\cdot t}(C,\lfloor \rho \cdot C \rfloor)}}{C}}\\
        &\geq \lim_{n \to \infty} \lim_{m \to \infty}{\frac{\log_2{M_{n\cdot t}(m \cdot n t, m \cdot n s)}}{m \cdot n t}}\\
        &\underset{\mathrm{Lemma~\ref{lemma:MqLowerBound}}}{\geq}
        \lim_{n \to \infty} \lim_{m \to \infty}{\frac{\log_2{\binom{nt}{ns}^m}}{m \cdot n t}}\\
        &= \lim_{n \to \infty} \frac{\log_2\binom{nt}{ns}}{nt} = \lim_{n \to \infty} \frac{\log_2\binom{nt}{\rho \cdot nt}}{nt}\hspace{-0.5ex} = \hspace{-0.5ex} H(\rho).
    \end{align*}
    Note that the above proof formally holds for $\rho \in \mathbb{Q}$.
    As for~(\ref{eq:Cap1}) it holds that
    \vspace{-0.5ex}
    \begin{align*}
        1 &= H\left(\frac{1}{2}\right) = \lim_{q \to \infty}\capsymbol(q, \frac{1}{2}) \leq \lim_{q \to \infty}\capsymbol(q)\\
        \vspace{-1ex}
        &\underset{\mathrm{Lemma~\ref{lemma:MqUpperBound}}}{\leq} \hspace{-1.5ex} \limsup_{C\rightarrow \infty}{\frac{\log_2{\left(\sum\limits_{L=0}^C \binom{C}{L}\right)}}{C}} 
        \hspace{-0.5ex} = \limsup_{C\rightarrow \infty}{\frac{\log_2{\left(2^C\right)}}{C}} \hspace{-0.5ex} = 1.
    \end{align*}
\end{proof}
\vspace{-5ex}
From~\cite{Lenz2021MultivariateAC}, it is known that
\begin{equation}\label{eq:andreasEq}
    \capsymbol(q,\rho)=
    \begin{cases}
    \rho\log_2{q}, 0 \leq \rho \leq \frac{2}{q+1}\\
    \rho\log_2\left(\sum_{i=1}^q x_q(\rho)^{i-\frac{1}{\rho}}\right),  \frac{2}{q+1} \hspace{-0.4ex}<\hspace{-0.4ex} \rho \hspace{-0.4ex}<\hspace{-0.4ex} 1,
    \end{cases}
\end{equation}
where $x_q(\rho)$ is the unique solution to $\sum_{i=1}^q(1-\rho i)x^i=0$ in the range $0 < x < 1$.
Furthermore, as per \cite{EFFICIENT9174272}, for any $q$ we have $\capsymbol(q) = -\log_2{x_q}$, where $x_q$, $(0< x_q <1)$ is the unique real solution of the polynomial $\sum_{i=1}^qx^i=1$.
Fig.~\ref{fig:Capacity_fixed_length} shows $\capsymbol(q,\rho)$ for some values of $q$ in relation to $H(\rho)$. Note that as $q$ grows, capacity is maximized for $\rho$ around $\frac{1}{2}$.
\subsection{Capacity Achieving Encoder}
Having explored the information-theoretical limits of photographic synthesis, our focus shifts to formulating an efficient encoder-decoder scheme capable of achieving the capacity.
For the special case of $\rho = \frac{2}{q+1}$, \cite{EFFICIENT9174272} presents an efficient encoder-decoder using a single redundancy symbol.
The time complexity of the encoder is linear and it encodes any string of length $L-1$ over an alphabet of size $q$ to a length-$L$ string over the same alphabet, such that its synthesis time is at most $C = \lfloor \frac{q+1}{2}\cdot L \rfloor$.
We demonstrate that this encoder's rate achieves the capacity.
The rate $R$ of this encoder satisfies
\begin{align*}
    R &= \lim_{L \to \infty}\frac{\log_2{q^{L-1}}}{\lfloor \frac{q+1}{2} \cdot L \rfloor} = \lim_{L \to \infty}\frac{(L-1)\cdot\log_2{q}}{\lfloor \frac{q+1}{2} \cdot L \rfloor} & \\
    &=\frac{2}{q+1}\cdot \log_2{q} = \rho\cdot \log_2{q} = \capsymbol(q,\rho=\frac{2}{q+1}).&
\end{align*}
For any other value of $\rho$ we want to achieve the best possible rate using a variation of this encoder. This problem is stated formally as follows.

\textbf{Problem 1.} Given $0 < \rho < 1$ and $q$ as the alphabet size to be used, our objective is to design a supersequence with an efficient encoder-decoder that achieves $\capsymbol(q,\rho)$.

Problem 1 is addressed in Section \ref{section:Encoders}.
\subsection{Optimizing The Synthesis Cost}
In this section we define a \emph{cost function} for photolithographic synthesis of symbols from $\Sigma_q$.
Let $\alpha, \beta \in \mathbb{R}^+$ be the cost of cycle synthesis and symbol synthesis, respectively. To derive a general cost function for storing $N$ bits using $C$ cycles, s.t. $\rho = \frac{L}{C}$ we are using the expression
\[
    \costsymbol(N,C,q,\rho) = \alpha C + \beta\cdot (\text{\#synthesized symbols}).
\]
Further, assume that we can achieve the capacity $\capsymbol(q,\rho)$ for any $q, \rho$, and so  we can thus store $C\cdot \capsymbol(q,\rho) = \frac{L}{\rho}\cdot\capsymbol(q,\rho)$ bits per oligo of length $L = \rho C$.
We determine the number of oligos to be synthesized to be $\frac{N}{\frac{L}{\rho}\cdot\capsymbol(q,\rho)}$ which implies that $\frac{\rho \cdot N}{\capsymbol(q,\rho)}$ symbols are synthesized.
Note that we should use the floor and ceil functions, but for the sake of simplicity, we have omitted them.
We therefore see that, under our assumptions, the optimal achievable cost is given by the formula
\begin{equation}\label{eq:costDefinition}
    \costsymbol^*(N,C,q,\rho) \triangleq \alpha C + \beta N \cdot \frac{\rho}{\capsymbol(q,\rho)}.
\end{equation}
\textbf{Problem 2.}
Our objective is to find the values of 
\begin{enumerate}
    \item $\text{arg}\min\limits_{\rho\in (0,1)}{\costsymbol^*(N, C, q, \rho)}$, for fixed $N,C,q$.
    \item $\text{arg}\min\limits_{\rho\in (0,1), q\in \mathbb{N}}{\costsymbol^*(N, C, q, \rho)}$, for fixed $N,C$.
\end{enumerate}
\begin{remark}
\small
Changes in $\rho$ will also impact the oligo length $L$, consequently altering the number of oligos to be synthesized.
This will also affect the barcode lengths required for indexing the oligos, influencing the overall cost.
While we acknowledge this effect, we do not address its specific implications.
\end{remark}
\section{Capacity Achieving Encoder}\label{section:Encoders}
While the capacity of photolithographic synthesis was calculated in prior work, an efficient encoder-decoder scheme that achieves the capacity, was only described, to our knowledge, for $\rho = \frac{2}{q+1}$.
We now address Problem 1, by introducing two families of efficient encoders applicable for a wider range of $\rho$ values, and achieving higher information rate.
\subsection{Lookup Table}\label{section:lookupTables}
\begin{figure}[H]
    \centerline{\includegraphics[width=7.5cm,height=4.5cm]{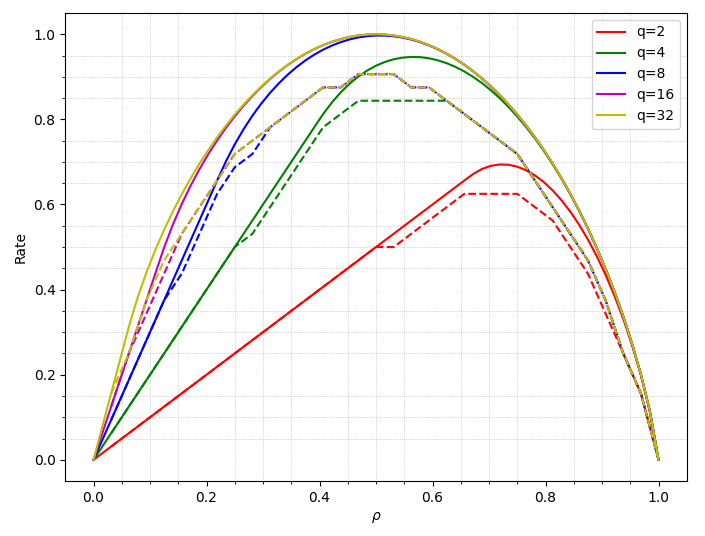}}
    \setlength{\abovecaptionskip}{-5pt}
    \caption{Information rates achievable by the construction of Section \ref{section:lookupTables}, compared to the related capacities $\capsymbol(q,\rho)$, for selected values of $q$.
    The dashed line represents the maximum achievable rate of the encoder, while the solid line represents the capacity. The value of $d$ for every pair of $(q,\rho)$ was chosen to be the largest integer s.t. $B\leq 32$.}
    \label{fig:CapacityFixedLength_MyEncoder_NoBorders}
\end{figure}

We start with a naive encoding approach by using a lookup table.
Let $d \in \mathbb{N}, \rho \in (0,1)$ s.t. $\rho \cdot d \cdot q \in \mathbb{N}$ and denote $L = \rho \cdot d \cdot q, C = d\cdot q$.
For $B = \lfloor \log_2\left(M_q(C, L) \right) \rfloor$, we pre-calculate a lookup table of $2^B$ entries, while each entry corresponds to a distinct subsequence of length $L$ out of $A_q[d\cdot q]$.
The encoding process iteratively encodes every $B$ bits into a string of length $L$ over $A_q[d\cdot q]$ using the lookup table.
We attain that the rate $R$ of this encoder satisfies $R = \frac{\lfloor \log_2\left(M_q(dq, \rho \cdot d \cdot q)\right) \rfloor}{dq}$.
Note that $\lim\limits_{d \to \infty} \frac{\lfloor \log_2\left(M_q(dq, \rho \cdot d \cdot q)\right) \rfloor}{dq} = \capsymbol(q, \rho)$.
Two major drawbacks of this approach is that the lookup table requires a substantial use of memory, and its calculation is time consuming.
Fig.~\ref{fig:CapacityFixedLength_MyEncoder_NoBorders} depicts the rate of this encoder using a lookup table of up to $2^{32}$ entries alongside with the capacity.
\subsection{Multi-size Alphabet Encoder}\label{section:MultiSizeAlphabetEncoder}
In this section we present a variation of the encoder presented in \cite{EFFICIENT9174272} (described above) where we use multiple alphabet sizes.
Initially, we analyze a scenario where a fraction of the synthesized oligos is encoded over an alphabet of size $q_1$, and the remainder is encoded over an alphabet of size $q_2$.
Using $C_1$ cycles over the alternating sequence $A_{q_1}$ and $C_2$ cycles over $A_{q_2}$ produces the machine's supersequence $A_{q_1}[C_1]\circ A_{q_2}[C_2]$,  which concatenates the first $C_1$ symbols of $A_{q_1}$ with the first $C_2$ symbols of $A_{q_2}$.
First we consider a case where half of the synthesized oligos is over $q_1$ and the other half is over $q_2$. In this case, to synthesize oligos of length $L$ it holds that $C_1 = \frac{q_1+1}{2}\cdot \frac{L}{2}$, $C_2 = \frac{q_2+1}{2}\cdot \frac{L}{2}$, $C = C_1+C_2$, $\rho = \frac{L}{C} = \frac{4}{q_1+q_2+2}$.
The rate $R$ of this encoder satisfies
\begin{small}
\begin{align*}
    R &= \lim_{L \to \infty}\hspace{-1ex} \frac{\log_2{\left(q_1^{\frac{L}{2}-1} \hspace{-2ex}\cdot q_2^{\frac{L}{2}-1}\right)}}{\frac{q_1+1}{2} \frac{L}{2} + \frac{q_2+1}{2} \frac{L}{2}} \hspace{-0.5ex}=\hspace{-0.5ex} \lim_{L \to \infty}\hspace{-1.25ex} \frac{(\frac{L}{2}\hspace{-0.5ex}-\hspace{-0.5ex}1)\log_2{q_1} \hspace{-0.5ex}+\hspace{-0.5ex} (\frac{L}{2}\hspace{-0.5ex}-\hspace{-0.5ex}1)\log_2{q_2}}{\frac{L}{4} \cdot (q_1+q_2+2)} \\
    &= \frac{4 \cdot (\frac{1}{2}\log_2{q_1} + \frac{1}{2}\log_2{q_2})}{q_1+q_2+2} = \rho \cdot (\frac{1}{2}\log_2{q_1} + \frac{1}{2}\log_2{q_2}).
\end{align*}
\end{small}
A more general approach is when using all $q$ alphabet sizes $1, 2, \dots ,q$, where each alphabet size $i$ is used on a fraction $\alpha_i$ of the synthesized string ($\sum_{i=1}^q \alpha_i = 1$).
Similarly to the previous case we get that $C = \sum_{i=1}^q C_i = \sum_{i=1}^q \frac{i + 1}{2}\cdot \alpha_i L$, $\rho = \frac{L}{C} = \frac{2}{1 + \sum_{i=1}^q i\alpha_i}$.
The rate $R$ of this encoder satisfies
\begin{align*}
    R &= \hspace{-0.5ex} \lim_{L \to \infty} \frac{\log_2{\left(\prod_{i=1}^q i^{\alpha_i L - 1}\right)}}{\sum_{i=1}^q \frac{i + 1}{2}\cdot \alpha_i L}
    =\hspace{-0.5ex} \lim_{L \to \infty} \frac{\sum_{i=1}^q (\alpha_i L - 1)\log_2{i}}{\sum_{i=1}^q \frac{i + 1}{2}\cdot \alpha_i L}\\
    &= \frac{2\sum_{i=1}^q \alpha_i \log_2{i}}{1+\sum_{i=1}^q i\alpha_i}
    = \rho \cdot \sum_{i=1}^q \alpha_i \log_2{i}.
\end{align*}
Consider a fixed $\rho \in (0,1)$, our goal is to maximize the rate $R$ that can be achieved for this $\rho$.
We also assume a fixed $q$ but we can control $\vec{\mathbf{\alpha}} \triangleq (\alpha_1, \alpha_2, \dots, \alpha_q)$.
Also recall that relevant values of $\vec{\mathbf{\alpha}}$ satisfy $\sum_{i=1}^q i\alpha_i = \frac{2}{\rho} -1$.

This yields the following linear programming (LP) system.
\begin{align*}
    &\max_{\Vec{\alpha}} \{\mathbf{c}^T\Vec{\alpha}\} 
    \textbf{ s.t. } \mathbf{A}\Vec{\alpha} = \mathbf{b}, \Vec{\alpha} \geq 0
\end{align*}
\begin{align*}
    &A = \begin{pmatrix}
        1 & 1 & \dots & 1 \\
        1 & 2 & \dots & q
    \end{pmatrix},
    \mathbf{b} = \begin{pmatrix}
        1\\
        \frac{2}{\rho} - 1
    \end{pmatrix},\\
    &\mathbf{c}^T = (\log_2{1}, \log_2{2}, \dots, \log_2{q}).
\end{align*}
Denoting by $\Vec{\alpha}^*$ an optimal solution leads us to the following.
\begin{claim}
    The optimal solution $\Vec{\alpha}^*$ has no two positive entries, with indices that are more than one index apart (i.e. no index $i$ so that $\alpha_i > 0$ and so that for some $k \geq 2$ also $\alpha_{i+k} > 0$).
\end{claim}
\begin{proof}
    Assume by contradiction that $\alpha_i, \alpha_{i+k} > 0$ for some $i \in \{1,2, \dots, q-2\}$ and $k \geq 2$.
    Denote $\alpha_i = x$, $\alpha_{i+k} = y$, $\alpha_{i+1} = u$, $\alpha_{i+k-1} = v$.
    Assume $y \leq x$ (the opposite case is similar).
    Define $\Vec{\alpha}_{new} = (\alpha_1', \alpha_2' \dots, \alpha_q')$ by $\alpha_i' = x - y$, $\alpha_{i+1}' = u + y$, $\alpha_{i+k-1}' = v + y$, $\alpha_{i+k}' = 0$, and $\forall j \neq i, i+1, i+k-1, i+k: \alpha_j' = \alpha_j$.
   Considering only the changed indices $i, i+1, i+k-1, i+k$, it can be shown that the constraints still hold for $\Vec{\alpha}_{new}$.
    As for the target function, let $r =  x\log_2(i) + u\log_2(i+1) + v\log_2(i+k-1) + y\log_2(i+k)$ and $r_{new} = (x-y)\log_2(i) + (u+y)\log_2(i+1) + (v+y)\log_2(i+k-1)$.
    We observe that the difference $r_{new} - r$ is
    \begin{align*}        
        &y(\log_2(i+1) + \log_2(i+k-1) - \log_2(i) - \log_2(i+k))\\
        &= y\log_2\frac{(i+1)(i+k-1)}{i(i+k)} 
        \hspace{-0.5ex} = y\log_2\frac{i^2+ik+k-1}{i^2+ik}\\
        &= y\log_2 (1 + \frac{k-1}{i^2+ik}),
    \end{align*}
    which is positive for all $i \geq 1, k\geq 2$.
    Therefore $r_{new} > r$ contradicting the optimality of $\Vec{\alpha}^*$.
\end{proof}
\begin{corollary}\label{cor:optIndex}
    For the optimal solution $\Vec{\alpha}^*$ there is an index $i \in \{1,2,\dots, q-1\}$ such that for all $j \neq i, i+1$,  $\alpha_j^* = 0$.
\end{corollary}
\begin{theorem}
    Let $s \in \{1,2,\dots, q-1\}$ and $\rho \in [\frac{2}{(s+1)+1}, \frac{2}{s+1}]$.
    The optimal rate $R^*$ for this configuration, obtained by the approach described above, is
    \[
        R^* = (\rho(s+2)-2)\log_2(s) + (2-\rho (s+1))\log_2(s+1).
    \]
\end{theorem}
\begin{proof}
    Given that $\rho \in [\frac{2}{(s+1)+1}, \frac{2}{s+1}]$ it holds that $\frac{2}{\rho}-1 \in [s, s+1]$.
    Let $i$ be the index of $\Vec{\alpha}^*$ from corollary~\ref{cor:optIndex}.
    From the first row of the matrix, it holds that $i\alpha^*_i + (i+1)\alpha^*_{i+1} \in [i, i+1]$.
    From the second row of the matrix it holds that $i\alpha^*_i + (i+1)\alpha^*_{i+1} \in [s, s+1]$.
    Therefore $i=s$.
    
    Since $\alpha^*_{s+1} = 1- \alpha^*_s$ we get $s\alpha^*_s + (s+1)(1- \alpha^*_s) = \frac{2}{\rho} -1$, yielding $\alpha_s^* = s+2-\frac{2}{\rho}$.
    Substituting\hspace{-0.3ex} into\hspace{-0.3ex} the\hspace{-0.3ex} rate function gives
    \begin{align*}
        R^* &= \rho \cdot (\alpha^*_s \log_2(s) + (1-\alpha^*_s)\log_2(s+1))\\
        &= \rho \cdot ((s+2-\frac{2}{\rho})\log_2(s) + (\frac{2}{\rho}-s-1)\log_2(s+1))\\
        &= (\rho(s+2)-2)\log_2(s) + (2-\rho (s+1))\log_2(s+1).
    \end{align*}
\end{proof}
\vspace{-3ex}
\subsection{Effective Encoder for $\rho = \frac{1}{2}$}\label{section:knuthEncoder}
In this section we address the specific case $\rho = \frac{1}{2}$.
In particular, we introduce an effective encoder-decoder for $\rho$ around $\frac{1}{2}$ applicable to all alphabet sizes.
Let $q$ be the alphabet size to be used, and let $f_q \triangleq \max \{n \in \mathbb{N}  : n + \lceil \log_2{n} \rceil + 1 \leq q\}$.
In our encoder we will use alphabet of size $q^* = f_q + \lceil \log_2(f_q) \rceil + 1$.
Note that $q^* \leq q$ so this is possible.
Our encoder achieves $\rho = \frac{\lfloor \frac{q^*}{2} \rfloor}{q^*}$ and rate $R = \frac{f_q}{q^*}$.

For $u \in \{0,1\}^*$ let $W_H(u)$ be the Hamming weight of $u$.
We define $I(u) \triangleq \{i : u_i = 1\}$, e.g. $I(0110) = \{2,3\}$.
Let $s$ be a string of the same length as $u$, we define $\pi(s,u)$, the projection of $s$ on $u$, to be obtained by keeping the indices $I(u)$ of $s$ and deleting the rest.
$\pi(s,u) \triangleq t_1t_2\dots t_{H(u)}$ such that $t_i = s_{I(u)_i}$, where $I(u)$ is sorted by ascending order.
For example, $\pi(abcd, 0110) = bc$.
Let $K(u)$ be the output of applying a variation of Knuth algorithm \cite{Knuth1986EfficientBC} using the redundancy symbols encoded with Gray code on $u$ \cite{Mambou2016EncodingAD}.
We first define the basic encoder $E_1: \{0,1\}^{f_q} \to \Sigma_{q^*}^{q^*}$ to be
\[
    E_1(u) \triangleq \pi(12 \dots q^*, K(u))
\]
For example, let $q=6$.
We then have $ f_q = 3$.
Consider $ u=100$.
We get that $K(u) = 010110$ and $E_1(u) = 245$.

Note that since $K$ is injective so is the encoder $E_1$.
\begin{claim}\label{claim:halfLengthBasicEncoder}
    For any $u \in \{0,1\}^{f_q}$, $E_1(u)$ is a subsequence of length $\lfloor \frac{q^*}{2} \rfloor$ of $A_{q^*}[q^*]$.
\end{claim}
Next, for all  $u = u_1u_2 \dots u_n$ such that $u_i \in \{0,1\}^{f_q}$, we define the encoder $E_2: \{0,1\}^{n\cdot f_q} \to \Sigma_{q^*}^{n q^*}$ to be
\[
    E_2(u) \triangleq E_1(u_1)E_1(u_2)\dots E_1(u_n).
\]
\begin{observation}\label{observ:E2_cycles}
    For any $u \in \{0,1\}^{n\cdot f_q}$, $E_2(u)$ is a subsequence of length $n \lfloor \frac{q^*}{2} \rfloor$ of $A_q[nq^*]$, and therefore can be synthesized with $nq^*$ cycles.
\end{observation}
\begin{theorem}\label{th:rate}
    For all $q \in \mathbb{N}$, the encoder $E_2$ achieves $\rho = \frac{\lfloor \frac{q^*}{2} \rfloor}{q^*}$ and rate $R = \frac{f_q}{q^*}$.
\end{theorem}
Theorem~\ref{th:rate} follows from Observation~\ref{observ:E2_cycles} and the fact that $R = \lim\limits_{n \to \infty} \frac{\log_2(2^{n\cdot f_q})}{n\cdot q^*} = \frac{f_q}{q^*}$.
Also note that
\[
    \lim\limits_{q \to \infty} \frac{f_q}{q^*} = \lim\limits_{n \to \infty} \frac{n}{n+ \lceil \log_2{n} \rceil +1} = 1 = \lim_{q \to \infty} \capsymbol(q,\frac{1}{2}).
\]
For $q = 4,8,16,32$, we get that the rate of this encoder is 
$0.5, 0.5714, 0.6875, 0.8125$ respectively, with the capacities $\capsymbol(q,0.5) = 0.92, 0.996, 0.99998, 1 - 10^{-9}$, respectively. 

Note that by allowing a flexible $\rho \leq \frac{1}{2}$, in each window of length $q$ we can encode any binary word of length $q-1$ using $1$ redundancy symbols.
For instance, if the weight of the binary word exceeds $\frac{q}{2}$, flipping the bits can be done, with the redundancy indicating whether the flipping occurred. The rate $R$ of this encoder would be $R = \frac{q-1}{q} = 1-\frac{1}{q}$.
\vspace{-2ex}
\subsection{Summary and Comparison Between the Approaches}\label{section:summaryAndComparison}
In Section \ref{section:lookupTables} we presented a naive encoding approach for a wide range of values of $\rho$ using a lookup table.
The rate of this encoding approaches the capacity as $d$ gets large.
But the calculation of the lookup table is time consuming and requires a substantial memory usage.

Next, in Section \ref{section:MultiSizeAlphabetEncoder} we presented a linear time complexity encoding scheme, which is a generalization of the encoding approach presented in \cite{EFFICIENT9174272}.
While this encoding achieves the capacity only for $\rho = \frac{2}{q+1}$, it is applicable for all $\rho \geq \frac{2}{q+1}$.

\begin{figure}[htbp]
    \centerline{\includegraphics[width=7.5cm,height=4.5cm]{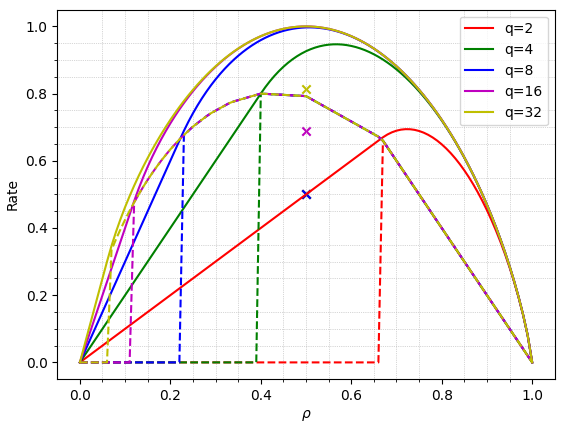}}
    \setlength{\abovecaptionskip}{-5pt}
    \caption{Information rates achievable by the construction of Section~\ref{section:MultiSizeAlphabetEncoder}, compared to the related capacities $\capsymbol(q,\rho)$, for selected values of $q$ and $\rho \in [0,1]$.
    Dashed lines represent the maximum achievable rate of the encoder of Section~\ref{section:MultiSizeAlphabetEncoder}, the x markers represent the rate achievable of the encoder of Section~\ref{section:knuthEncoder} and solid lines represent capacities.}
    \label{fig:Capacity_fixed_length_EitansEncoder}
\end{figure}

Finally, in Section \ref{section:knuthEncoder} we presented a linear time complexity encoding scheme for $\rho = \frac{1}{2}$ that approaches the capacity as $q$ gets large.
Although this scheme, strictly speaking, is applicable only for $\rho = \frac{1}{2}$ (or $\rho = \frac{1}{2} - \frac{1}{2q}$), note that values close to $\frac{1}{2}$ achieve the highest capacities. Fig.~\ref{fig:Capacity_fixed_length_EitansEncoder} depicts the capacity and the maximum achievable rate using the methods from Section \ref{section:MultiSizeAlphabetEncoder} and Section \ref{section:knuthEncoder}.
In summary, this section introduced two families of efficient encoders that achieve the capacity for certain values of $\rho$, while providing good performance in other configurations.

\section{Optimizing The Synthesis Cost}\label{section:CostOptimization}
\begin{figure}[htbp]
    \centerline{\includegraphics[width=7.5cm,height=4.5cm]{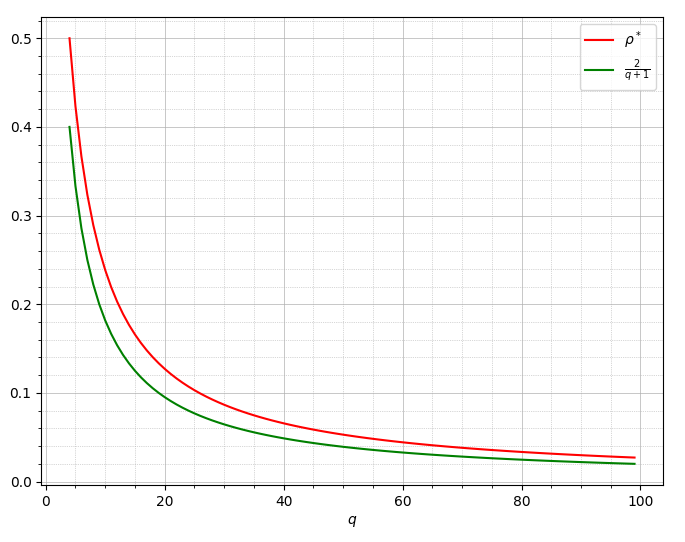}}
    \setlength{\abovecaptionskip}{-5pt}
    \caption{Values of $\frac{2}{q+1}$ and $\rho^*$, as a function of $q$. The space between the lines is the interval of interest, as proven in Theorem~\ref{theorem:IntervalOfInterest}.}
    \label{fig:RhoStarAndqOPT_noBorders}
\end{figure}

In this section we address Problem 2, seeking to identify the optimal parameters that minimize the synthesis cost as defined  above by $\costsymbol^*(N,C,q,\rho) = \alpha C + \beta N \cdot \frac{\rho}{\capsymbol(q,\rho)}$.
\begin{observation}
    $\underset{\rho\in [0,1]}{\arg\min}~{\costsymbol^*(N, C, q, \rho)}
    \hspace{-0.5ex}
    = \hspace{-0.5ex} \underset{\rho\in [0,1]}{\arg\min}{\frac{\rho}{\capsymbol(q, \rho)}}.$
\end{observation}
This follows directly from the definition of the cost function, assuming all variables except $\rho$ are fixed.
\begin{claim}\label{claim:lowerBoundRho}
    \hspace{-0.5ex}$\underset{\rho\in [0,1]}{\arg\min}~{\costsymbol^*\hspace{-0.5ex} (N, C, q, \rho)} 
    \hspace{-0.75ex}=\hspace{-0.5ex} \underset{\rho\in [\frac{2}{q+1},1)}{\arg\min}{\costsymbol^*\hspace{-0.5ex} (N, C, q, \rho)}.$
\end{claim}
\begin{proof}
    For all $\rho \in (0,\frac{2}{q+1}]$ it holds that $\capsymbol(q,\rho) = \rho\log_2{q}$ (see eq. (\ref{eq:andreasEq})). Hence,
    \[
        \costsymbol^*(N, C, q, \rho) \hspace{-0.5ex} = \hspace{-0.5ex} \alpha \cdot C + \beta \cdot N \frac{\rho}{\capsymbol(q,\rho)} \hspace{-0.5ex}=\hspace{-0.5ex} \alpha \cdot C + \beta \cdot \frac{N}{\log_2{q}}.
    \]
Since the cost function is independent of $\rho$ within this interval, we obtain the result.
\end{proof}
\begin{claim}\label{claim:bestQ}
    \hspace{-1ex}$\min\limits_{\rho\in (0,1)}\hspace{-1ex}{\costsymbol^*(N, C, q, \rho)} 
    >\hspace{-1ex} \min\limits_{\rho\in (0,1)}\hspace{-1ex}{\costsymbol^*(N, C, q+1, \rho)}.$
\end{claim}
\begin{observation}\label{observation:capAndEntropy}
    For all $q \in \mathbb{N}, \rho \in (0,1):$ $\frac{\rho}{\capsymbol(q,\rho)} \geq \frac{\rho}{H(\rho)}$.
\end{observation}
\begin{lemma}\label{lemma:monEntropy}
    The function $f(\rho) = \frac{\rho}{H(\rho)}$ is monotonously increasing for all $\rho \in (0,1)$, and $\lim\limits_{\rho \to 0} f(\rho) = 0$.
\end{lemma}
\begin{theorem}\label{theorem:IntervalOfInterest}
    Let $\rho^* \in (0,1)$ be the solution to the equation $\frac{\rho}{H(\rho)} = \frac{1}{\log_2{q}}$.
    Then,
    \[
        \underset{\rho\in [0,1]}{\arg\min}{\frac{\rho}{\capsymbol(q, \rho)}}
        = \underset{\rho\in [\frac{2}{q+1},\rho^*]}{\arg\min}{\frac{\rho}{\capsymbol(q, \rho)}}.
    \]
\end{theorem}
\begin{proof}
    According to Claim~\ref{claim:lowerBoundRho} it is enough to consider only $\rho \in [\frac{2}{q+1}, 1)$.
    As for the right end of the interval, for all $\rho \geq \rho^*$
    \[
        \frac{\rho}{\capsymbol(q,\rho)} > \frac{\rho}{H(\rho)} \geq \frac{\rho^*}{H(\rho^*)}
        = \frac{1}{\log_2{q}} = \frac{\frac{2}{q+1}}{\capsymbol(q, \frac{2}{q+1})},
    \]
    where the first and second inequalities is by Observation~\ref{observation:capAndEntropy} and Lemma~\ref{lemma:monEntropy}.
    The last step follows from Equation~(\ref{eq:andreasEq}).
\end{proof}
\vspace{-1ex}
Fig. \ref{fig:RhoStarAndqOPT_noBorders} depicts $\rho^*$ and $\frac{2}{q+1}$ as a function of $q$.
Note that $\rho^*$ is an implicit quantity.
Using the monotonicity stated in Lemma~\ref{lemma:monEntropy} it can be easily approximated to any desirable degree.
Turning our attention to Problem 2.2, let $\costsymbol_{OPT}^*(N,C) \hspace{-0.5ex} \triangleq \hspace{-0.5ex} \inf \{\costsymbol^*(N,C,q,\rho)\hspace{-0.5ex} :\hspace{-0.5ex} \rho \hspace{-0.5ex} \in \hspace{-0.5ex} (0,1), q \hspace{-0.5ex} \in \hspace{-0.5ex} \mathbb{N}\}$.
\begin{claim}
It holds that $\costsymbol_{OPT}^*(N,C) = \alpha C$.
\end{claim}
\begin{proof}
From Claim \ref{claim:bestQ} we get
\begin{align*}
    \costsymbol_{OPT}^*(N,C)  &= \inf \{\lim_{q \to \infty} \costsymbol^*(N,C,q,\rho) : \rho \in (0,1)\}\\
    &= \inf \{\alpha C + \beta N\frac{\rho}{H(\rho)} : \rho \in (0,1)\}.
\end{align*}
From Lemma \ref{lemma:monEntropy} we get that
\[
    \inf \hspace{-0.5ex} \left\{\alpha C + \beta N\frac{\rho}{H(\rho)} : \rho \in (0,1)\right\}
   \hspace{-1ex} = \hspace{-0.5ex} \lim_{\rho \to 0}{\alpha C \hspace{-0.5ex} + \hspace{-0.5ex} \beta N\frac{\rho}{H(\rho)}}
    \hspace{-0.5ex} = \hspace{-0.5ex} \alpha C.
\]
\end{proof}
\vspace{-2ex}
When the alphabet size is bounded by $Q$ we get
\begin{align*}
    \costsymbol_{OPT}^*(N,C) &= \inf \left\{\costsymbol^*(N,C,q,\rho) \hspace{-0.5ex}:\hspace{-0.5ex} \rho \in (0,1), q \leq Q \right\}\\
    &= \inf \left\{\costsymbol^*(N,C,Q,\rho) : \rho \in (0,1)\right\}\\
    &\in \hspace{-0.5ex} \left\{ \alpha C + \beta N \frac{\rho}{\capsymbol(Q, \rho)} \hspace{-0.5ex} : \rho \in [\frac{2}{Q+1}, \rho^*]\right\},
\end{align*}
assuming a characterization based on the implicit quantity $\rho^*$.

\section{Discussion}\label{section:Discussion}
We study computational aspects of photolithographic synthesis.
We start by expanding on literature results and presenting efficient coding and decoding at values of $\rho$ that were not previously addressed. The number $\rho \in [0,1]$ represents the fraction of  synthesis cycles when using a total of $C$ cycles to synthesize oligos of length $L < C$. Previous work calculated the exact capacity at a single value of $\rho$, for every alphabet size. This value, however, does not yield the most efficient scheme. We expand on these results, presenting effective coding and show, in particular, that for large enough values of $q$ (that can be achieved, e.g. by composite alphabets \cite{Anavy2019DataSI, Preuss2024}), $\rho = \frac{1}{2}$, or close to it, is the most interesting value, achieving rates that approach capacity.   
We defined cost using~(\ref{eq:costDefinition}).
We note two key considerations.
First, this cost is calculated assuming that capacity is achieved, and therefore is denoted $\costsymbol^*$. Further investigation is needed in order to estimate the cost at lower rates.
Second, while $\beta$ is assumed constant in this paper, larger $q$  will result in higher cost per every added nucleotide.
Therefore, it is natural to formulate our questions in terms of $\beta$ as a function of $q$.
This would require a deeper understanding of $\beta$ in relation to $q$, which is currently lacking.
We hope to address this question in future work.
We note again that $\rho^* = \rho^*(q)$, an important parameter for the development presented in this paper and that $\rho^*$ is implicit.
Nonetheless, it can be well approximated thanks to the monotonicity of $\frac{\rho}{H(\rho)}$ in $(0,1)$ (see Section \ref{section:CostOptimization}).
Finally - we note that efficient coding for values of $q$ that are not large can be obtained by a lookup table, using the method described in Section \ref{section:Encoders}.
\section*{Acknowledgment}
\small
We thank the Yaakobi and Yakhini research groups as well as many DiDAX members for useful discussions.
\newpage
\bibliographystyle{abbrv}
\bibliography{Ref}

\begin{thebibliography}{10}

\bibitem{Anavy2019DataSI}
L.~Anavy, I.~Vaknin, O.~Atar, R.~Amit, and Z.~Yakhini.
\newblock Data storage in {DNA} with fewer synthesis cycles using composite {DNA} letters.
\newblock {\em Nature Biotechnology}, 37:1229 -- 1236, 2019.

\bibitem{Antkowiak2020LowCD}
P.~L. Antkowiak, J.~Lietard, M.~Z. Darestani, M.~M. Somoza, W.~J. Stark, R.~Heckel, and R.~N. Grass.
\newblock Low cost {DNA} data storage using photolithographic synthesis and advanced information reconstruction and error correction.
\newblock {\em Nature Communications}, 11, 2020.

\bibitem{behr2024}
J.~Behr, T.~Michel, M.~Giridhar, S.~Santhosh, A.~Das, H.~Sabzalipoor, et~al.
\newblock An open-source advanced maskless synthesizer for light-directed chemical synthesis of large nucleic acid libraries and microarrays.
\newblock {\em ChemRxiv}, 2024.
\newblock This content is a preprint and has not been peer-reviewed.

\bibitem{BLAWAT20161011}
M.~Blawat, K.~Gaedke, I.~Hütter, X.-M. Chen, B.~Turczyk, S.~Inverso, B.~W. Pruitt, and G.~M. Church.
\newblock Forward error correction for dna data storage.
\newblock {\em Procedia Computer Science}, 80:1011--1022, 2016.
\newblock International Conference on Computational Science 2016, ICCS 2016, 6-8 June 2016, San Diego, California, USA.

\bibitem{Choi2019HighIC}
Y.~Choi, T.~Ryu, A.~C. Lee, H.~Choi, H.-B. Lee, J.~Park, S.-H. Song, S.~Kim, H.~Kim, W.~Park, and S.~Kwon.
\newblock High information capacity dna-based data storage with augmented encoding characters using degenerate bases.
\newblock {\em Scientific Reports}, 9, 2019.

\bibitem{Church}
G.~Church, Y.~Gao, and S.~Kosuri.
\newblock Next-generation digital information storage in {DNA}.
\newblock {\em Science (New York, N.Y.)}, 337:1628, 08 2012.

\bibitem{US-20210141568-A1}
A.~L. et~al.
\newblock Molecular data storage systems and methods, 2021.
\newblock US Patent Application.

\bibitem{US-20230040158-A1}
A.~L. et~al.
\newblock Molecular data storage systems and methods, 2023.
\newblock US Patent Application.

\bibitem{US-20240013063-A1}
R.~N. et~al.
\newblock Systems for nucleic acid-based data storage, 2024.
\newblock US Patent Application.

\bibitem{Goldman}
N.~Goldman, P.~Bertone, S.~Chen, C.~Dessimoz, E.~M. LeProust, B.~Sipos, and E.~Birney.
\newblock Towards practical, high-capacity, low-maintenance information storage in synthesized {DNA}.
\newblock {\em Nature}, 494(7435):77--80, 2013.

\bibitem{Grass2015RobustCP}
R.~N. Grass, R.~Heckel, M.~Puddu, D.~Paunescu, and W.~J. Stark.
\newblock Robust chemical preservation of digital information on dna in silica with error-correcting codes.
\newblock {\em Angewandte Chemie}, 54 8:2552--5, 2015.

\bibitem{Grass_2015}
R.~N. Grass, R.~Heckel, M.~Puddu, D.~Paunescu, and W.~J. Stark.
\newblock Robust chemical preservation of digital information on {DNA} in silica with error-correcting codes.
\newblock {\em Angewandte Chemie International Edition}, 54(8):2552--2555, feb 2015.

\bibitem{Hirschberg2000TightBO}
D.~S. Hirschberg and M.~R{\'e}gnier.
\newblock Tight bounds on the number of string subsequences daniel s.
\newblock {\em Journal of Discrete Algorithms}, 2000.

\bibitem{Knuth1986EfficientBC}
D.~E. Knuth.
\newblock Efficient balanced codes.
\newblock {\em IEEE Trans. Inf. Theory}, 32:51--53, 1986.

\bibitem{EFFICIENT9174272}
A.~Lenz, Y.~Liu, C.~Rashtchian, P.~H. Siegel, A.~Wachter-Zeh, and E.~Yaakobi.
\newblock Coding for efficient dna synthesis.
\newblock In {\em 2020 IEEE International Symposium on Information Theory (ISIT)}, pages 2885--2890, 2020.

\bibitem{Lenz2021MultivariateAC}
A.~Lenz, S.~Melczer, C.~Rashtchian, and P.~H. Siegel.
\newblock Multivariate analytic combinatorics for cost constrained channels and subsequence enumeration.
\newblock {\em ArXiv}, abs/2111.06105, 2021.

\bibitem{Lietard2018HighDensityRM}
J.~Lietard, D.~Ameur, M.~J. Damha, and M.~M. Somoza.
\newblock High‐density rna microarrays synthesized in situ by photolithography.
\newblock {\em Angewandte Chemie (International Ed. in English)}, 57:15257 -- 15261, 2018.

\bibitem{Mambou2016EncodingAD}
E.~N. Mambou and T.~G. Swart.
\newblock Encoding and decoding of balanced q-ary sequences using a gray code prefix.
\newblock {\em 2016 IEEE International Symposium on Information Theory (ISIT)}, pages 380--384, 2016.

\bibitem{Preuss2024}
I.~Preuss, M.~Rosenberg, Z.~Yakhini, and L.~Anavy.
\newblock Efficient dna-based data storage using shortmer combinatorial encoding.
\newblock {\em Scientific Reports}, 14(1):7731, Apr 2024.

\bibitem{Schaudy2023EnzymaticSO}
E.~Schaudy, J.~Lietard, and M.~M. Somoza.
\newblock Enzymatic synthesis of high‐density rna microarrays.
\newblock {\em Current Protocols}, 3, 2023.

\bibitem{Shafir2021SequenceRU}
R.~Shafir, O.~Sabary, L.~Anavy, E.~Yaakobi, and Z.~Yakhini.
\newblock Sequence reconstruction under stutter noise in enzymatic dna synthesis.
\newblock {\em 2021 IEEE Information Theory Workshop (ITW)}, pages 1--6, 2021.

\bibitem{Shafir2024SequenceDA}
R.~Shafir, O.~Sabary, L.~Anavy, E.~Yaakobi, and Z.~Yakhini.
\newblock Sequence design and reconstruction under the repeat channel in enzymatic dna synthesis.
\newblock {\em IEEE Transactions on Communications}, 72:675--691, 2024.

\bibitem{Yan2023}
Y.~Yan, N.~Pinnamaneni, S.~Chalapati, C.~Crosbie, and R.~Appuswamy.
\newblock Scaling logical density of dna storage with enzymatically-ligated composite motifs.
\newblock {\em Scientific Reports}, 13(1):15978, Sep 2023.

\end{thebibliography}
\vspace{12pt}
\end{document}